\documentclass[showpacs,twocolumn,floatfix,superscriptaddress]{revtex4}
\usepackage{graphicx}

\newcommand{\bq}{\begin{equation}}
\newcommand{\ee}{\end{equation}}
\newcommand{\fr}[2]{\frac{#1}{#2}}
\newcommand{\eps}{\varepsilon}

\begin{document}

\draft

\title{ The Enhanced Sensitivity of the Transmission Phase of a
Quantum Dot  to Kondo Correlations}
\author{P. G. Silvestrov}
\affiliation{Instituut-Lorentz, Universiteit Leiden, P.O. Box 9506, 2300 RA
Leiden, The Netherlands}
\affiliation{Budker Institute of Nuclear Physics, 630090 Novosibirsk, Russia}
\author{Y. Imry}
\affiliation{Weizmann Institute of science, Rehovot 76100, Israel}


\date{17 December 2001}


\begin{abstract}

The strong sensitivity  of the transmission phase through a
quantum dot embedded into one arm of a two-wave Aharonov-Bohm
interferometer to the Kondo effect is explained. The enhancement
takes place because of the buildup of the exchange scattering on
the dot due to Kondo correlations even much above $T_{K}$. The
enhanced exchange competes with the potential scattering, which is
always weak. Both cases of the Anderson impurity model and a
multilevel quantum dot are considered. In the latter case in
addition to the description of peculiar phase behavior a mechanism
leading to ferromagnetic Kondo coupling in quantum dots is
proposed.

\end{abstract}

\maketitle


{\bf Introduction.} Among the
observations~\cite{Kondo1,Kondo2,Kondo3} of the Kondo effect in
quantum dots(QD)~\cite{KondoT}, two experiments~\cite{Yang1,Yang2}
were devoted to the measurement of the phase shift of the
transmitted electron. These experiments were aimed at a direct
observation of one of the most fundamental
predictions\cite{Langreth,Nozieres} of the Kondo model: the $\phi=
\pi/2$ phase shift experienced by the scattered electron after the
spin of the impurity is screened into a singlet.

In agreement with the
numerical renormalization group calculations\cite{Delft}
for an Anderson impurity\cite{Anderson}
in an Aharonov-Bohm(AB) interferometer,
the development of a plateau of $\phi$
in the Kondo valley was seen in ref.~\cite{Yang1}.
Unexpected however was the strong sensitivity of the phase
to Kondo correlations observed in the experiment: the phase
saturated at $\phi\approx const$, while the
conductance in the valley was always
well below the unitary limit, indicating the absence of Kondo screening.

Our first aim in this paper will be just to explain this strong
Kondo effect in the phase. We will see
that although the nontrivial phase behavior is indeed governed by
the Kondo physics, the main changes of phase take place at
temperatures parametrically larger than $T_K$.
In a sense, we show that the phase changes in the regime where,
although the spin of the impurity is not screened, the running
Kondo coupling constant exceeds parametrically its bare value.
Based on this result we will be able to describe the phase in the
same regime in the more complicated case of multilevel QD. 
In addition, in case of many levels 
we found a mechanism leading to the alternation of sign of the Kondo
coupling from antiferromagnetic to ferromagnetic within one
Coulomb blockade valley.
This effect should be useful for interpretation
of the conductance structures
observed in recent experiments~\cite{chess}.

The transmission phase is measured by embedding the quantum dot
into one arm of the two-wave AB interferometer~\cite{Heiblum}. Let
$A_{s S}^{d}$ be the transmission amplitude for an electron with
spin projection $s$ through the QD having a spin projection $S$.
Respectively, let $A_{s}^r$ be the transmission amplitude through
the second reference arm. (Obviously, the spin- flip processes
occurring in the QD and not in the reference arm do not contribute
to the interference.) Now, the part of the current oscillating
with the change of the magnetic flux threading the interferometer
takes the form \bq\label{AB} G_{AB} \propto Re \sum_{s
S}A_{s}^{r*}A_{s S}^{d} = Re A^{r*}\sum_{s S}A_{s S}^{d} \ . \ee
We use the fact that (for the weak magnetic field used in the
experiment) the transmission through the reference arm does not
depend on spin. Corrections to eq.~(\ref{AB}) due to multiple
scattering in the interferometer are suppressed in the open
geometry used in the experiment~\cite{Heiblum,Yang1,Yang2}. By
measuring the relative phase of the AB oscillations at different
values of the gate voltage one extracts the information about the
transmission phase. We see that in fact in the interference
experiment not a single amplitude is measured, but the sum of all
possible amplitudes corresponding to different orientations of the
spins of both the transmitted electron and the QD~\cite{Oreg}.

{\bf The single-level Anderson model.}
Far from the charging resonances the interaction of the lead
electrons with the dot is described by the Kondo Hamiltonian
\bq\label{HKondo}
H_K= \sum_{iks} \eps_k c^{i\dagger}_{ks}
c^i_{ks}
 + \sum_{ij} [J^{ij}_0\hat{\vec \sigma}^{ij}_e{\vec S}_d
+V^{ij}_0\hat{n}^{ij}_e] \ .
\ee
Here $i$ and $j$ denote left(L)
and right(R) leads, the operator $c^i_{ks}$ annihilates
the electron in lead $i$ with momentum $k$ and spin $s$. The
Pauli operators and the density of the conduction-electrons
on the impurity are given by
\bq\label{density}
\hat{\vec \sigma}^{ij}_e=\sum_{k k's
s'}c^{i\dagger}_{ks} {\vec \sigma}_{ss'}c^j_{k's'} \ , \
\hat{n}^{ij}_e=\sum_{k k's}c^{i\dagger}_{ks} c^j_{k's} \ .
\ee
Explicit formulas for $J_0$ and $V_0$ will be found below from the
tunnelling Hamiltonian describing the QD. We will
consider
a time-reversal symmetric system. Then
the matrices $J^{ij},V^{ij}$ are real and symmetric.

The interaction of the conduction electron with the spin of the
dot may be diagonalized by an orthogonal
transformation~\cite{Pustilnik} of $c_L,c_R$ into new
operators $c_u$ and $c_v$, described by the angle $\theta$, $\tan
2\theta={2J_0^{LR}}/{(J_0^{LL}-J_0^{RR})}$. This gives
\bq\label{HKuv}
H_{\sigma S}= (J_0^u\hat{\vec \sigma}_u+
J_0^v\hat{\vec \sigma}_v){\vec S}_d \ .
\ee
As long as $J_0^u,
J_0^v\ll 1$ the two couplings are renormalized independently
\bq\label{RGuv}
1/J^{u,v}=1/J_0^{u,v} +
4\nu\ln\left( {T}/{\Gamma} \right).
\ee
Here $\nu = \nu_L=\nu_R$ is the density
of states in the leads. In the case of
antiferromagnetic coupling ($J > 0$ ) this formula may be written
in the form ${4\nu}J=1/{\ln(T/T_K)}$ with $T_K$ being the
Kondo temperature. Crucial for the understanding of the phase
behavior is that in the leading order only the spin-dependent part
of the Hamiltonian (\ref{HKondo}) is renormalized, while the
scalar coupling remains unchanged,  $V=V_0=const$.

In the simplest case of only one level in the dot (Anderson
impurity model), only one mode
$c=(t_Lc_L +t_Rc_R)/t$
is coupled to the dot via the tunnelling matrix element
$t=\sqrt{|t_L|^2+|t_R|^2}$, while the second mode remains
completely decoupled. The bare values of the coupling constants in
the Kondo Hamiltonian (\ref{HKondo}) are now given by the second
order of perturbation theory
\bq\label{bare}
\fr{t^2}{-\eps_d}=V_0+\fr{J_0}{2}  \ ; \
\fr{t^2}{-(U+\eps_d)}=V_0-\fr{J_0}{2} .
\ee
Here $\eps_d<0$ is the
energy of the impurity level, $U$ is the charging energy
and the Fermi energy
is $E_F=0$.

The
non spin-flip
transmission  amplitudes for parallel and
antiparallel spins of the dot and the electron are
\begin{eqnarray}\label{Aren}
&&A^d_{\uparrow\uparrow}=\fr{t_L^\dagger t_R}{-\eps_d}
\left(1
-\fr{i\Gamma}{-2\eps_d}\right) \ , \\
&&A^d_{\uparrow\downarrow} =
\fr{t_L^\dagger t_R}{-(U+\eps_d)}
\left(1
-\fr{i\Gamma}{-2(U+\eps_d)}\right) \ , \nonumber
\end{eqnarray}
where $\Gamma=2\pi t^2\nu$. The imaginary part of the amplitudes
here is formally of the fourth order in the tunnelling amplitudes
$t_{L,R}$. However, the calculation of the phase of transmission
amplitude requires only the $S$-matrix for non-spin-flip
scattering to second order in $t$. That is why the Kondo
correlations do not contribute to the phase in the leading order
and  eq.~(\ref{Aren}) coincides with the expansion of the usual
Breit-Wigner formula. The Kondo effect appears in the real part of
the amplitudes at the order $\sim t^4$, which we will take into
account via the renormalization~(\ref{RGuv}). In order to find the
AB current one should rewrite the sum of the two amplitudes
$A^d_{\uparrow\uparrow}$ and $A^d_{\uparrow\downarrow}$
in terms of the scalar and the magnetic couplings~(\ref{bare}) and then
replace $J_0$ by $J$~(\ref{RGuv}), which gives 
\bq\label{result}
G_{AB}\propto Re A^{r*} [ V_0-i\pi\nu (V_0^2+{J^2}/{4})] \ . 
\ee
This formula is the central result of the paper. It should be
compared with a usual conductance of a Kondo quantum dot, obtained
by adding $|A_{\uparrow\uparrow}|^2$,
$|A_{\uparrow\downarrow}|^2$~(\ref{Aren}) and the corresponding
spin-flip contribution, 
\bq\label{Gusual}
G=2\fr{e^2}{h}\Gamma^2\fr{|t_L t_R|^2}{t^8} \left(
V_0^2+\fr{3}{2}J^2\right) . 
\ee 
Remember that $V_0$ is not
renormalized, while $J\sim 1/\ln(\Gamma/T_K)$. Eqs. (\ref{Gusual})
and (\ref{result}) are justified only for $\nu J\ll 1$. This does
not allow for the quantitative description of the conductance in
the unitary limit, where $G\approx 2e^2/h$. On the other hand, the
phase shift close to $\pi /2$ develops at \bq \nu J_0\ll \nu
J\sim\sqrt{\nu V_0}\ll 1. \ee This gives the temperature scale,
explaining the high sensitivity of the transmission phase to the
Kondo effect observed in the experiment~\cite{Yang1},
$\ln(T/T_K)\sim \sqrt{\ln(\Gamma/T_K)}$.

\begin{figure}[t]
\includegraphics[width=8.3cm]{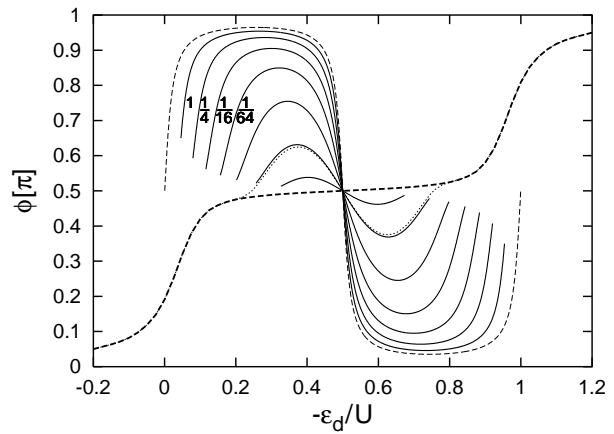}
\caption{ The phase
$\phi$ as a function of the depth of the impurity level
$-\eps_d$ for $\Gamma=U/30$. Solid lines show the
calculated $\phi(\eps_d)$ for $T=\Gamma, \Gamma/4, ...
,\Gamma /4^6 $. Since the theory is valid only for $T\gg T_K$,
the curves are shown only for $T>2 T_K(\eps_d)$. The dotted line
depicts schematically  the expected $\phi(\eps_d)$ for 
$T=\Gamma/1024$. The thin dashed line shows the phase for the sum of two
Breit-Wigner resonances and the thick dashed line is the 
conventional Bethe-ansatz 
solution for $T\equiv 0$.}
\end{figure}

Fig.~1 shows the phase $\phi(\eps_d)$ found via eq.~(\ref{result})
for $\Gamma=U/30$ and
different temperatures. Due to a node of $V_0$ at $\eps_d=-U/2$
the phase equals exactly $\pi/2$ in the middle of the valley. The
width of the phase drop at higher temperature ($T\sim \Gamma$) is
$\sim\Gamma$. The exact 
Bethe-ansatz solution~\cite{Bethe} for $T\ll T_K$ yields the broad 
plateau between resonances with $\phi\approx \pi/2$~\cite{Delft}.
However, due to a
strong dependence of $T_K$ on the position of the impurity level
$\eps_d$ the phase changes from $\phi=0,\pi$ to $\phi=\pi/2$ in a
rather nonuniform way. 
For intermediate temperatures the phase first develops a
shoulder with $\phi\approx\pi/2$ ($T\ll T_K(\eps_d)$), 
while in the center~($T\gg T_K(\eps_d)$) 
pronounced maximum and minimum are formed to the left and right 
of the point $\eps_d=-U/2$.
This structure may be seen on the experimental figures of
ref.~\cite{Yang1}, but appears  not to have been noticed in the
numerical calculation of ref.~\cite{Delft}.
The conductance (\ref{Gusual}) in the middle of the valley
for the same parameters of the dot as in Fig.~1 varies from
$G=0.013 e^2/h$
for $J=J_0$ to
$G=0.42 e^2/h$
for the lowest
temperature $T=\Gamma/4^6$. We see that the phase behavior is
well developed when the conductance is
sufficiently below the unitary limit.

{\bf
Multilevel QD.}
The Anderson
model is not sufficient  to
explain all features of the phase behavior observed in the
experiments~\cite{Yang1,Yang2}.
One possibility~\cite{exchange}
to go beyond this model is to include the exchange interaction,
allowing for higher spin of the multilevel dot.
We will now generalize the foregoing discussion to  an arbitrary
spin of the dot. The measured AB amplitude
is now given by
\bq\label{ampAB}
\sum_S(V_0 +S J )^{LR} [1-i\pi\nu \sum_{j=L,R} (V_0 +S J )^{jj}] .
\ee
Here we averaged over the $z$-projection of the dot spin $S$.
The renormalized couplings are given by
$J^{LR}=(J^v-J^u)\sin 2\theta$, and $J^{LL}+J^{RR}=J^u+J^v$.

The next step is the derivation of $J_0$ and $V_0$
from the microscopic model, generalizing the single-level
result of Eq.~(\ref{bare}).
Let the dot have two levels $a$ and
$b$. The ground state energy for the dot containing one, two and
three electrons is $\eps_a$, $\eps_{ab}^1$ and $\eps_{aab}$
respectively. The lower index shows which levels are occupied.
In case of two electrons we show also the total spin ($0$ or $1$).
This structure of the energy
levels may be achieved if for example $\eps_a< \eps_b$ and
$\eps_{aa}=2\eps_{a}+U$, $\eps_{bb}=2\eps_{b}+U$,
$\eps_{ab}^1=\eps_a+\eps_b+U-U_e$,
$\eps_{ab}^0=\eps_a+\eps_b+U+U_e$, where $U$ and $U_e$ are the
direct and exchange Coulomb interactions.
We assume similar sensitivity of single-electron levels to the gate voltage
$d\eps_{a,b}/dV_g=-1$.
Level crossings lead to
charging resonances at $\eps_a=0$, $\eps_{ab}^1=\eps_a$, etc.
Remarkably, the Kondo effect and phase behavior in such a QD
has a very unusual form even for the $S=1/2$ valley.

In the first valley ($S={1}/{2}$) the transition amplitudes
$e\!\!\uparrow \!d\!\!\uparrow \rightarrow e\!\!\uparrow
\!d\!\!\uparrow$ and $e\!\!\uparrow \!d\!\!\downarrow \rightarrow
e\!\!\uparrow \!d\!\!\downarrow$ give the bare scalar and magnetic
coupling constants
\begin{eqnarray}\label{upupdown}
V_0^{ij}+\fr{J_0^{ij}}{2}=\fr{t_a^{i}t_a^j}{-\eps_a}
&+&\fr{t_b^{i}t_b^j}{\eps_a-\eps_{ab}^1} \\
V_0^{ij}-\fr{J_0^{ij}}{2}=\fr{t_a^{i}t_a^j}
{\eps_a-\eps_{aa}^0}
&+&\fr{1}{2}\left(
\fr{t_b^{i}t_b^j}{\eps_a-\eps_{ab}^1}+
\fr{t_b^{i}t_b^j}{\eps_a-\eps_{ab}^0}\right)  .
\nonumber
\end{eqnarray}
The amplitudes $e\!\!\uparrow \!d+
\rightarrow e\!\!\uparrow \!d+$ and $e\!\!\uparrow \!d-
\rightarrow e\!\!\uparrow \!d-$ in the
second valley ($S=1$) are
\begin{eqnarray}\label{up+-}
&&V_0^{ij}+J_0^{ij}=\fr{t_a^{i}t_a^j}{\eps_b-\eps_{ab}^1}
+\fr{t_b^{i}t_b^j}{\eps_a-\eps_{ab}^1}
\\
&&V_0^{ij}-J_0^{ij}=\fr{t_a^{i}t_a^j}{\eps_{ab}^1
-\eps_{aab}}+
\fr{t_b^{i}t_b^j}{\eps_{ab}^1-\eps_{abb}} . \nonumber
\end{eqnarray}
The third valley ($S = 1/2$) is considered similarly.
After the bare coupling constants are
found, a straightforward calculation leads to the diagonal Kondo
Hamiltonian~(\ref{HKuv}) which is renormalized via~(\ref{RGuv}).

In the vicinity of the second resonance the
term
$\sim {t_b^{i}t_b^j}/{(\eps_a-\eps_{ab}^1)}$
dominates in Eqs.~(\ref{upupdown},\ref{up+-}),
leading to $|J^u_0|\gg |J^v_0|$ here.
The sign of the large coupling is always
positive, $J^u_0>0$, in the $S=1$ valley (antiferromagnetic
interaction), and always negative, $J^u_0<0$, in the $S=1/2$
valley (ferromagnetic interaction).
This result may be used for
explanation of the chess-board structure observed
in the $V_g-B$ (gate voltage -- magnetic field) plane
in Kondo experiments\cite{chess}.
Consider the line on the plane corresponding to one charging
resonance. With the change of $B$ on the
odd side of the line the spin of the QD jumps between $S=0$ and $S=1$
due to the crossing of energy levels in the dot. The Kondo effect
is present only for the $S=1$ squares. However, as we found from
Eqs.~(\ref{upupdown},\ref{up+-}), on the odd side of the
same resonance the Kondo effect will show up at the
$1/2\rightarrow 0$ transition, but is always suppressed
at the $1/2\rightarrow 1$ transition.

\begin{figure}[t]
\includegraphics[width=8.3cm]{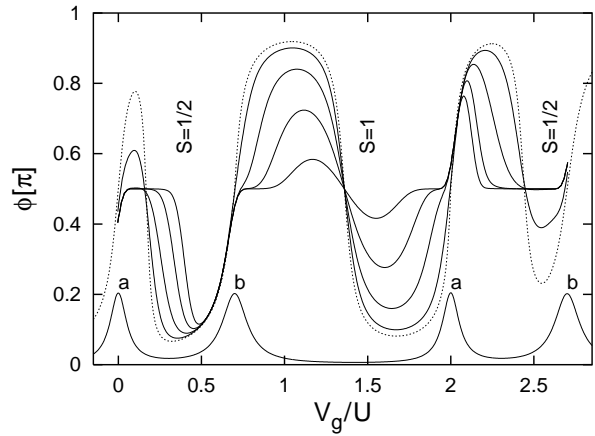}
\caption{ The phase $\phi(V_g)$ for a sequence of valleys $S=1/2,
1, 1/2$. Dotted line: Breit-Wigner. Solid lines: the phase at
different temperatures $T=\Gamma, \Gamma/8,...$.
The bottom line shows the schematic conductance of such a dot,
letters $a,b$ indicate, to which level the electron is added.
}
\end{figure}

Eq.~(\ref{ampAB}) allows us to find the transmission phase. For
real tunnelling amplitudes $t^i_k$ there may be only two different
phase behaviors depending on the relative sign of $t^L_a t^R_a$ and
$t^L_b t^R_b$. Fig.~2 shows the phase $\phi(V_g)$ when $t^L_a
t^R_a$ and $t^L_b t^R_b$ have the same sign.
The phase at high
temperatures for this choice of signs increases by $\pi$ at each
resonance and decreases by $\pi$ in the valleys.
Moreover, to simplify  the case shown in Fig.~2 further, we choose
$t^L_a t^R_b=t^R_a t^L_b$ (this includes an ``experimentally
desirable'' situation of symmetric coupling $t^L_i=t^R_i$). In
this particular case
one of the constants $J^{u,v}$ vanishes exactly.
The phase in the $S=1$ (middle) valley in Fig.~2
behaves similarly to the phase in the Anderson
model (Fig.~1). In Fig.~2 the $S=1/2$ valleys have an $S=0$ valley
on one side and an $S=1$ one on the other side. The usual Kondo
effect occurs only on one half of the $S=1/2$ valley boarding on
the $S=0$. Thus with decreasing temperature the conductance {\it
increases} and {\it decreases} in different portions of one and
the same valley. The phase for $S=1/2$ approaches $\phi\approx
\pi/2$ on the antiferromagnetic side of the valley, but remains
$\phi\approx 0,\pi$ on the ferromagnetic side. At the transition
from antiferromagnetic to ferromagnetic behaviors the phase drops,
but only by $\Delta\phi\approx\pi/2$. Phase drops by only a
fraction of $\pi$ are seen in Fig.~2 of the recent experimental
paper~\cite{Yang2}.

For opposite signs of $t_a^Lt_a^R$ and $t_b^Lt_b^R$
the phase at high temperatures increases by $\pi$ at each resonance
and stays close to $\#\pi$ in all three valleys (no phase drops).
In the Kondo regime ($T\rightarrow 0$) the phase in the $S=1/2$ valleys
approaches the plateaus $\phi\approx \pi/2$ and $\phi\approx 7\pi/2$.
In the $S=1$ valley the phase at
$T\rightarrow T_K$ for the considered signs of tunnelling
amplitudes first forms a $3\pi/2$ plateau in the left
half of the valley, then increases quickly by $+\pi$ and stays
at $5\pi/2$ in the right half of the valley. This $+\pi$ jump
of phase in the Kondo regime takes place due to vanishing
of the transmission coupling $J_0^{LR}=0$ at a certain point
in the $S=1$ valley. Notice however that the
experiments~\cite{Heiblum,Yang1,Yang2} always reported  the
existence of a phase drop (and never increase of phase) in the
Coulomb blockade valley. Theoretically only a finite
sequence of phase drops may be explained (see~\cite{SI}).

We now turn to discuss the  case where the  spin on the dot is
zero. The AB current Eq.~(\ref{AB}) is proportional to the single
amplitude $A^d_{s0}\equiv A^d_{-s0}$ in this case.
Since poles of the amplitude, $A^d$, correspond to charging
resonances, the only reason for change of phase in the valley may
be a node, $A^d\sim \eps-\eps_{0}$. The zeroes of the amplitude
do not occur in one dimensional case, but
should be found in about a half  of
the spin zero valleys in two-dimensional QD.
The transmitted electron wave
function is $\psi=A^de^{ikx}$ and a vanishing of the amplitude at
$\eps=\eps_{0}$ means the existence of solution of the
Schr{\"o}dinger equation with $\psi\equiv0$ in one lead. If the
Hamiltonian is time-reversal symmetric, the complex conjugated
wave function should also be a solution with the same boundary
condition and $\eps=\eps_{0}^*$, hence $Im\eps_0\equiv 0$. The
phase changes abruptly by $\pi$ when $\eps_0$ crosses the Fermi
energy.
The transition acquires a finite width due to
thermal excitations 
$\delta
V\sim\Gamma e^{-{\Delta}/{T}}$,
where $\Delta$ is the 
level spacing in the dot.
The sign of
$\pi$-jump for $S_d=0$ is negative for symmetric coupling, but it
may be positive, $+\pi$, in the asymmetric case.

To summarize, we explained in this paper the extra-strong
sensitivity of the transmission phase through a quantum dot to the
Kondo correlations.
For the Anderson impurity model we found new features of $\phi$,
which were not noticed in existing numerical simulations. 
Even more complicated phase behavior is predicted for a multilevel QD.
In
addition, we propose an explanation of the chess-board structure
of the Kondo effect observed in the conductance of a QD in several
experiments~\cite{chess}. Finally, we show that the phase jumps in spin zero
valleys have a vanishing width at zero temperature,
a feature which
needs an  experimental verification.

We acknowledge  valuable discussions with N.~Andrei, A.~Georges,
L.~I.~Glazman and M.~Pustilnik.
This project was supported by a Center of Excellence of the Israel Science
Foundation, Jerusalem, by the German-Israeli Foundation (GIF),
by the  German Federal Ministry of Education and Research (BMBF)
within
the Framework of the German-Israeli Project Cooperation (DIP)
and
by the Chaires Internationales Blaisse Pascal. The work of P.G.S.
was supported by Dutch Science Foundation NWO/FOM. The work was
concluded at the Institute for Theoretical Physics at the
University of California Santa Barbara, supported by
the National Science Foundation under Grant No.~PHY99-07949.



\begin{thebibliography}{99}
\vspace{-1cm}


\bibitem{Kondo1} D.~Goldhaber-Gordon {\it et al.}, Nature {\bf 391},
156 (1998).

\bibitem{Kondo2} S.~M.~Cronenwett {\it et al.},
Science {\bf 281}, 540 (1998).

\bibitem{Kondo3} J.~Schmid {\it et al.},
Physica B {\bf 256-258}, 182
(1998).

\bibitem{KondoT}
L.~I.~Glazman and M.~E.~Raikh, JETP Lett. {\bf 47},
452 (1988);
T.~K.~Ng and P.~A.~Lee, Phys. Rev. Lett. {\bf 61}, 1768
(1988).
N.~S.~Wingreen and Y.~Meir, Phys.~Rev. {\bf B 49}, 11040
(1994).

\bibitem{Yang1} Y.~Ji {\it et al.},
Science, {\bf 290}, 779 (2000).

\bibitem{Yang2} Y.~Ji, M.~Heiblum, H.~Shtrikman, 
Phys. Rev. Lett. {\bf 88}, 076601 (2002).

\bibitem{Langreth} D. C. Langreth, Phys. Rev. {\bf 150}, 516 (1966).

\bibitem{Nozieres} P. Nozi\`{e}res. J. Low Temp. Phys. {\bf 17}, 31
(1974).

\bibitem{Delft} U.~Gerland {\it et al.}, Phys. Rev. Lett. {\bf 84},
3710 (2000).

\bibitem{Anderson} P.~W.~Anderson, Phys. Rev. {\bf 124}, 41 (1961).

\bibitem{chess} J. Schmid {\it et al.}, Phys. Rev. Lett. {\bf 84},
5824 (2000); D.~Sprinzak {\it et al.}, cond-mat/0109402.


\bibitem{Heiblum}
R.~Schuster {\it et al.}, Nature, {\bf 385}, 417 (1997). 
For a
discussion of the conditions for a two-path interference, see: O.
Entin-Wohlman {\it et al.}, Phys. Rev. Lett. {\bf 88}, 166801 (2002).

\bibitem{Oreg} Y.~Oreg and Y.~Gefen, Phys. Rev. {\bf B 55},
13726 (1997).

\bibitem{Pustilnik}  M. Pustilnik and L. I. Glazman,
Phys. Rev. Lett. {\bf 87},
216601 (2001).

\bibitem{Bethe} P.~B.~Wiegmann and A.~M.~Tsvelick, J.~Phys.~C: 
Solid State Phys., {\bf 16}, 2281 (1983).

\bibitem{exchange} P.~W.~Brouwer {\it et al.},
Phys.~Rev. {\bf B 60}, R13977 (1999);
H.~U.~Baranger {\it et al.},
Phys.~Rev. {\bf B 61}, R2425 (2000);
P.~Jacquod, A.~D.~Stone, Phys. Rev. Lett. {\bf 84}, 3938 (2000);
I.~L.~Kurland {\it et al.},
Phys. Rev. Lett. {\bf 86}, 3380 (2001).


\bibitem{SI} P. G. Silvestrov and Y. Imry, Phys. Rev. Lett.
{\bf 85}, 2565 (2000).



\end{thebibliography}
\end{document}